\documentclass[12pt]{article}
\usepackage{amssymb,amsmath,epsfig}

\begin{document}

\title{\bf  Gravitational Perfect Fluid Collapse in Gauss-Bonnet Gravity}
\author{G. Abbas \thanks{ghulamabbas@iub.edu.pk} and  M. Tahir \thanks{tmuhammad0064@gmail.com}
\\Department of Mathematics, The Islamia University \\of Bahawalpur, Bahawalpur-63100, Pakistan.}
\date{}
\maketitle 
\begin{abstract}
The Einstein Gauss-Bonnet theory of gravity is the low energy limit of heterotic super-symmetric string theory. This paper deals gravitational collapse of perfect fluid in Einstein Gauss-Bonnet gravity by considering the Lemaitre - Tolman - Bondi metric. For this purpose, the closed form of exact solution of equations of motion has been determined by using the conservation of stress-energy tensor and the condition of marginally bound shells. It has been investigated that the presence of Gauss-Bonnet coupling term $\alpha>0$ and pressure of the fluid modifies the structure and time formation of singularity. In this analysis singularity form earlier than horizon, so end state of the collapse is a naked singularity depending on the initial data. But this singularity is weak and timelike that goes against the investigation of general relativity. \textbf{Some future research directions are mentioned at the end of the paper}.
\end{abstract}
\section{Introduction}

In recent years a great interest has been developed for the study of higher order curvature theories of gravity \cite{5}-\cite{8}. Among all these higher curvature theories of gravity, the most widely studied theory of gravity is known as Einstein-Gauss-Bonnet (EGB) gravity. The Lagrangian of EGB gravity is the particular form of Lagrangian of Lovelock theory \cite{9}. In 4-dimensional vacuum theory of gravity, the most general Lovelock theory is a combination of the $0^{th}$ and $1^{st}$ Euler density, in other words, general relativity (GR) with a cosmological constant. Also, it has been observed that the $2^{nd}$ Euler density is known as the Gauss-Bonnet combination, which is is topological invariant in 4-dimensions and does not contribute to the dynamical equations of the motion if included in the action \cite{10}. Further, a simple and natural way to make GB combination dynamical in 4-dimensional theory is to couple it to a dynamical scalar field. The perturbation method has been used to see the effects of GB coupling term $\alpha>0$ on the dynamical instability of non-static anisotropic fluid spheres \cite{11}. The thermal aspects of gravitating source in $5D$ EGB theory of gravity have been explored in detail by coupling the heat transport equation with the dynamical equations \cite{12}. Boulware and Deser \cite{8} have formulated the static black hole exact solutions in greater than or equal to five dimensional theories of gravity with a four dimensional GB term modifying the usual Einstein-Hilbert action \cite{14}.

Several LTB-like solutions Einstein field equations have been explored in higher order theories of gravity, which include the higher derivative curvature terms in their action. The higher curvature correction to action have a great effect on the topological structure of singularity appearing during the gravitational collapse of massive star \cite{Ch1}. In EGB theory of gravity with $\alpha>0$, there exists a massive, naked and un-central singularity only in $5D$, while such singularity is disallowed in $D\geq6$ \cite{Ch2}. It has been the subject of great interest for many theoretical physicists to explore the slowly rotating BH solutions in Gauss-Bonnet theory of gravity. However, due to nonlinearity of field equations in GB gravity, it is very hard to obtain the exact
analytic rotating black hole solutions in the framework of this theory. Therefore by introducing a small angular momentum as a perturbation into a rotating system, some BH solutions have investigated in the past \cite{Ch3, Ch33}. Also, such slowly rotating BHs solutions exist in the EGB theory with AdS term \cite{Ch4}. Using the linear order of the rotation parameter $\textit{a}$, the slowly rotating charged/uncharged BHs solutions in AdS third order Lovelock gravity have been investigated in \cite{Ch5}, which have some interesting physical features.

During the recent decade AdS BHs and especially their thermodynamics
have attracted the attention of many researchers due to the AdS/CFT duality. In the AdS EGB gravity, the thermodynamical relations such temperature and entropy
of the charged BHs get no corrections from rotation parameter $\textit{a}$ \cite{Ch6}. Zou et al.\cite{Ch7} have explored the thermodynamics of GB-Born-Infeld BHs in AdS space. Also, they have calculated the mass, temperature, entropy and heat of the resulting BHs.  In third order Lovelock AdS gravity, the thermodynamic and conserved quantities of BHs with flat horizons are independent of Lovelock coefficients \cite{Ch8}. Such interests in higher order theories of gravity have motivated us to study the gravitational collapse of perfect fluid in EGB with LTB model.

The stars are composed of some nuclear matter and gravitational collapse is the phenomenon in which the stars are continuously gravitating and attracted towards their centers due to the gravitational interaction of its particles. According to the singularity theorem \cite{29}, the space-time singularities are generated during the continual gravitational collapse of massive stars. During the recent decades, it has been an interesting problem in astronomy and astrophysics to determine the final fate of gravitational collapse . In this connection, Oppenheimer and Snyder \cite{30} are the pioneer who found the black hole as end state of the dust collapse by using the static Schwarzschild spacetime as an exterior spacetime and a Friedmann like solution as an interior spacetim. Later on this work was extended with positive and negative cosmological constant \cite{31,32}. Several authors \cite{10a}-\cite{36} have discussed the phenomena of gravitational collapse using the dissipative and viscous fluid in general relativity. The lack of the analytical consequences, leads to the unproven conjectures namely, cosmic censorship conjecture (CCC) \cite{40}, hoop conjecture (HC)\cite{41}, and Seifert's conjecture \cite{42}. Oppenheimer and Snyder \cite{30}, considered a homogeneous spherical star with zero rotation and vanishing internal pressure in these ideal conditions, the cloud collapse simultaneously to a spacetime singularity, which is enclosed by an event horizon. Further, it is interesting to study the gravitational collapse of stars with some realistic matter and geometry.

Recently, Jhingan and Ghosh \cite{44}, have studied the dust spherical collapse in five dimensions with GB term. They find the exact solutions in closed form with the marginally bound conditions. In this paper, we extended the work of Jhingan and Ghosh \cite{44}, to case of perfect fluid with marginally bound conditions. The paper has been arranged as follows: In section 2, we present the exact solution of field equation. Section \textbf{3} deals with singularity analysis. In the last section, we summaries the results of the paper.

\section{Exact Solution of Field Equations in Einstein Gauss-Bonnet Gravity}
Here, the required $5D$ gravitational action is
\begin{eqnarray}\label{ac1}
S = \int d^5 x \sqrt {-g} [\frac{1}{2\kappa^2_{5}}(R+\alpha L_{GB})]+S_{matter},
\end{eqnarray}
here $R$ and $\kappa_{5} \equiv \sqrt{8 \pi G_{5}} $ , $\alpha$ are Ricci scalar, gravitational constant in $5D$, and Gauss-Bonnet coupling constant, respectively. In this case Gauss-Bonnet Lagrangian is
\begin{eqnarray}\label{a11}
 L_{GB} = R^2 -4R_{\mu\nu} R^{\mu\nu} +R_{\mu\nu\gamma\lambda} R^{\mu\nu\gamma\lambda},
\end{eqnarray}
  The above action appears as the low energy limit of heterotic super-string theory \cite{8} and $\alpha$ is the inverse string tension which is usually taken as positive finite, so we restrict ourselves to the case when $\alpha\geq 0 $. The variation of action (\ref{ac1}) yields the following form of equations of motion in Einstein Gauss-Bonnet gravity
\begin{eqnarray}\label{a111}
G_{\mu\nu} + \alpha H_{\mu\nu} = T_{\mu\nu},
\end{eqnarray}
where $G_{\mu\nu}=R_{\mu\nu} - \frac{1}{2} g_{\mu\nu}R$
is the Einstein tensor and
\begin{eqnarray}\label{a1111}
 H_{\mu\nu}=2[RR_{\mu\nu} -2 R_{\mu \alpha} R^{\alpha}_{\nu} -2R^{\alpha \beta} R_{\mu\alpha\nu \beta}+R^{\alpha \beta \gamma}_{\mu} R_{\nu\alpha \beta \gamma}] -\frac{1}{2} g_{\mu\nu} L_{GB}
\end{eqnarray}
is the Lanczos tensor.
 The stress-energy tensor for perfect fluid is
\begin{eqnarray}\label{t1}
      T_{\mu\nu}=\Big( \rho(r,t) +p(r,t)\Big)V_{\mu}V_{\nu}+p g_{\mu\nu} ,
\end{eqnarray}
where $V_{\mu}= \delta^{t}_{\mu}$ is $5D$ velocity, $\rho(r,t)$ is energy density and $p(r,t)$ is isotropic pressure due to fluid distribution in the interior region of a star. The LTB metric with co-moving coordinates in $5D$ case \cite{44, 46, 47, 48} is
\begin{eqnarray}\label{met1}
 ds^2 = -dt^2 +B^2 dr^2 + C^2 d\Omega^2_{3},
\end{eqnarray}
where $B=B(r,t)$ and $C=C(r,t)$, and $d\Omega^2_{3} = (d\theta^2 +\sin^2\theta (d\phi^2+\sin^2\phi d\psi^2))$.\\

 The set of independent field equations in Einstein Gauss-Bonnet gravity for the metric (\ref{met1}) and stress-energy tensor (\ref{t1}) are
\begin{eqnarray}\nonumber
 &&\frac{12(C'^2 -B^2 (1+\dot{C}^2))}{C^3 B^5}\Bigg(C' B' +B^2 \dot{C} \dot{B} -B C''\Bigg)\alpha
 - \frac{3}{B^3 C^2}\Bigg(B^3 (1+ \dot{C}^2)\\\label{f1}&&+B^2C \dot{C} \dot{B} + C C' B' -B(CC'' +C'^2)\Bigg) = -\rho,\\\label{f2}
&&-12 \alpha \Bigg(\frac{1}{C^3} -\frac{C'^2}{B^2C^3}+\frac{\dot{C}^2}{C^3}\Bigg)\ddot{C}+3\frac{C'^2}{B^2C^2}-\frac{3\Bigg(1+ \dot{C}^2+C \ddot{C}\Bigg)}{C^2} = p,\\\nonumber
&&\frac{4 \alpha}{B^4 C^2}\Bigg[-2B\Bigg(B'C'+B^2\dot{B} \dot{C}-BC''\Bigg)\ddot{C}+B\Bigg(C'^2 -B^2 (1+\dot{C}^2)\Bigg)\ddot{B}\\\nonumber&&+2\Bigg(\dot{B}C'-B\dot{C}'\Bigg)\Bigg]-\frac{1}{B^3C^2}\Bigg[B^3\Bigg(1+\dot{C}^2+2C\ddot{C}\Bigg)+B^2C\Bigg(2\dot{C}\dot{B}+C\ddot{B}\Bigg)\\\label{f3}&&+2CC'B'-2B\Bigg(CC''+C'^2\Bigg)\Bigg] = p,
\end{eqnarray}
\begin{eqnarray}\nonumber
\label{f01}
&&\frac{12 \alpha}{B^5C^3}\Bigg(\dot{B}C'-B\dot{C}'\Bigg)\Bigg(B^2\Bigg(1+\dot{C}^2\Bigg)-C'^2\Bigg)-3 \frac{B \dot{C}'-\dot{B}C'}{B^3C} =0,
\end{eqnarray}
 where, $. =\partial_t$ and $'=\partial_r$. After some simplification Eq.(\ref{f01}), yields two families of solutions in the following form

\begin{eqnarray}\label{s1}
 B(t,r) &&= \frac{C'}{Z},\\\label{s2}
 B(t,r) &&= \pm \frac{2\sqrt{\alpha}C'}{\left(C^2 + 4 \alpha (\dot{C}^2 +1)\right)^{1/2}},
\end{eqnarray}
where $Z=Z(r)$ is function of integration. The solution for $B(r,t)$ in Eq.(\ref{s1}) is similar to 5D-LTB solution \cite{46, 47, 48}, while solution in Eq.(\ref{s2}) is trivial for $\alpha \rightarrow 0$, hence we take non-trivial form of $B(r,t)$ given in Eq.(\ref{s1}). Now Eq.(\ref{f2}) with Eq.(\ref{s1}) gives
\begin{eqnarray}\label{fi}
 \frac{\ddot{C}}{C}=\frac{\dot{C}^2 - (Z^2-1)+\frac{C^2p}{3}}{4 \alpha(Z^2 -1 -\dot{C}^2)-C^2}.
\end{eqnarray}
Now we have to solve above equation analytically for $C(r,t)$, this requires that we have to integrate the above
twice with respect to $t$. Since this equation involves the unknown function $p(r,t)$, when we try to integrate Eq.(\ref{fi}), we can not get the exact solution because there is unknown function $p(r,t)$, to get ride of it, we try to make it \textit{at least} independent of $t$. To this end, we apply the conservation of energy-momentum tensor which make it independent of $r$. But our purpose is to make is independent of $t$, for this we take $p(t)$ as a polynomial in $t$, then after some simplification, we make it constant (as taken in \cite{D1}). This whole procedure is explained as follows

The conservation of energy-momentum tensor gives
\begin{equation}\label{p0}
\frac{\partial p}{\partial r}=0,~~~~~\Rightarrow p=p(t).
\end{equation}
We consider $p$ as a polynomial in $t$ as given by \cite{D1}
\begin{equation}\label{44b}
p(t)=p_0(\frac{t}{T})^{-c},
\end{equation}
where $T$ is the constant time introduced in the problem due to
physical reason by re-scaling of $t,~p_0$ and $c$ are positive
constants. Further, for the integration of Eq.(\ref{fi}), we
take $c=0$ in Eq.(\ref{44b}), so that
\begin{equation}\label{aa}
p(t)=p_0.
\end{equation}
There are many other choices for $p(t)$ , so Eq.(\ref{44b}) is not always a unique choice for p(t). For example it can be treated with $c\neq0$, but such choices do not provide the results in closed form which reduce to $5D$ dimensional perfect fluid collapse in the limit $\alpha\rightarrow0$ \cite{36,47,48}, and not recover the results of $5D$ dust collapse in Einstein Gauss-Bonnet gravity as $p\rightarrow0$ \cite{44}. That is why we have taken the pressure as constant which is a better choice in the present situation. Further, some other choice of $p(t)$ as defined in Ref.\cite{D1,D2}, may produce interesting numerical results but such solutions may be considered explicitly in future investigation by taking anisotropic fluid.

Using above equation in Eq.(\ref{fi}) and intergrading, one get
\begin{eqnarray}\label{fv}
\dot{C}^2[1-4 \alpha (\frac{Z^2 -1}{C^2})]=(Z^2 -1)+\frac{\zeta}{C^2}-\frac{1}{6}C^2p_0-2 \alpha \frac{\dot{C}^4}{C^2}.
\end{eqnarray}
where $\zeta=\zeta(r)$, is integration function and assumed as a mass function. The equation (\ref{fv}) is the main equation of the system.  By using Eqs. (\ref{s1})and (\ref{fv}) into Eq.(\ref{f1}), we get
\begin{eqnarray}\label{as1}
 \zeta' =\frac{2}{3}C^3C' (\rho +p_0).
\end{eqnarray}
The integration of above equation yields
\begin{eqnarray}\label{a1}
 \zeta(r)=\frac{2}{3}\int (\rho +p_0)C^3 dC,
\end{eqnarray}
 Here, we have used $\zeta(0)=0$.
 The validity of energy conditions is necessary a requirement
for a physically reasonable energy-momentum tensor. The energy conditions
for perfect fluid in this case are following
\begin{eqnarray}
{NEC}:\quad&&\rho+p_0\geq0,\\
{WEC}:\quad&&\rho\geq0, \quad \rho+p_0\geq0,\\
{SEC}:\quad&&\rho+p_0\geq0, \quad \rho+3p_0\geq0.\\\nonumber
\end{eqnarray}

\begin{figure}
\begin{center}
\includegraphics[width=80mm]{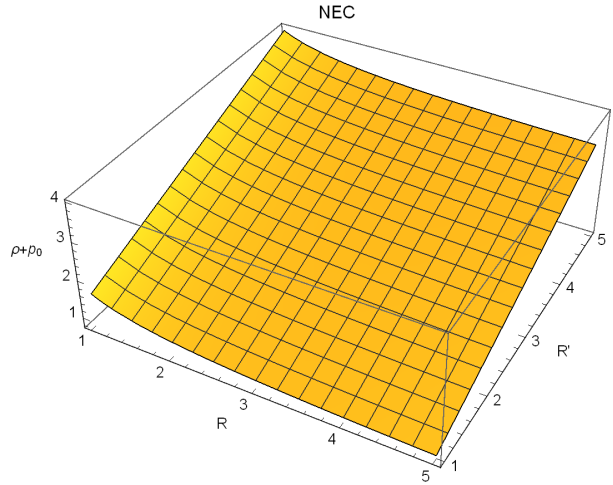}
\caption{This graph shows the validity of null energy condition for $p_0>0$, $\zeta'(r)>0$.}
\end{center}
\end{figure}
\begin{figure}
\begin{center}
\includegraphics[width=80mm]{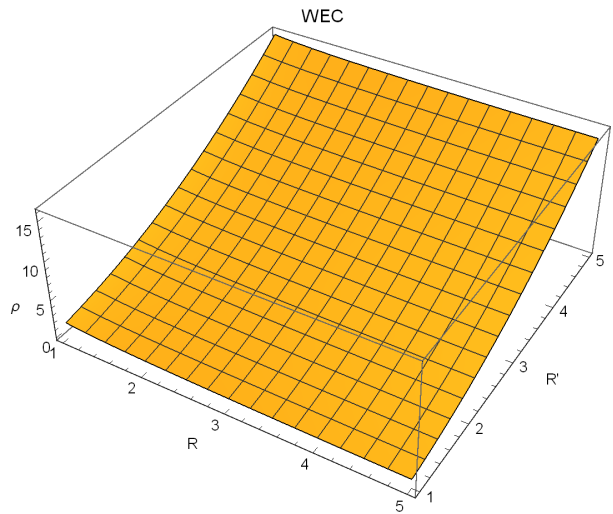}
\caption{This graph shows the validity of weak energy condition}.
\end{center}
\end{figure}
\begin{figure}
\begin{center}
\includegraphics[width=80mm]{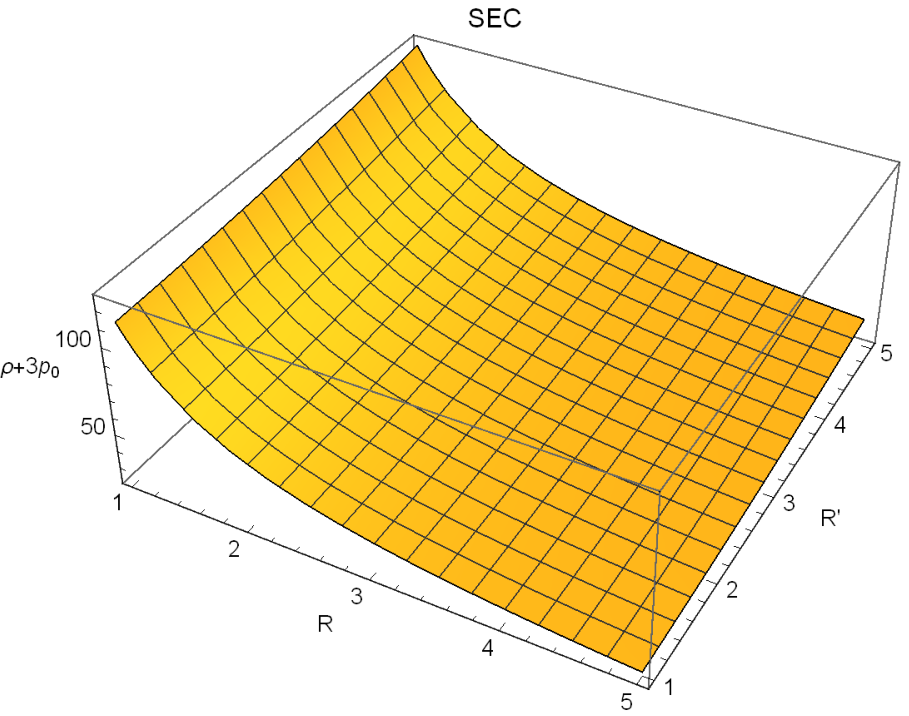}
\caption{This graph shows the validity of null energy condition for $p_0>0$.}
\end{center}
\end{figure}
From Eq.(\ref{as1}), we see that for $\zeta'(r)>0$,\quad$C>0$ and $C'>0$, one gets $\rho+p_0>0$ and null energy condition is valid as shown in \textbf{Fig.1}.
Using  Eq.(\ref{s2}), marginally bound condition $Z=1$, $\dot{C}<0$, $\dot{C'}<0$ along with above mentioned conditions in the field equations (\ref{f2}) and (\ref{s1}), we get $\rho\geq0$ as shown in \textbf{Fig.2}. Also, field equations (\ref{f2}), (\ref{s1}) and (\ref{aa}) along with the above mentioned conditions yield  $\rho+3p_0\geq0$ as shown in \textbf{Fig.3}. Hence null, weak and strong energy conditions are valid with restrictions on $C$ and its derivatives.

In the limit $\alpha \rightarrow 0$, Eq.(\ref{fv}) gives the $5D$ perfect fluid solutions \cite{47, 48}
\begin{eqnarray}\label{b1}
 \dot{C}^2 = Z^2 -1+\frac{\zeta}{C^2}-\frac{1}{6}C^2p_0.
\end{eqnarray}
 Now above equation yields three solutions, hyperbolic, parabolic, and elliptic solutions if $Z > 1$, $Z = 1$ or $Z< 1$, respectively \cite{47, 48}. For simplicity, we take $Z = 1$, which is marginally bound case.
 In order to proceed ahead, we write Eq.(\ref{fv}), in the following simplified form
\begin{eqnarray}\label{b2}
 \dot{C}^2 = (Z^2 -1)- \frac{C^2}{4 \alpha}\Big(1\mp \sqrt{\left[1 + \frac{16 \alpha^2}{C^4}(Z^2 -1)^2 -\frac{4}{3} \alpha p_0+\frac{8 \alpha \zeta}{C^4}\right]}\Big),
\end{eqnarray}
corresponding to $\mp$ sign in Eq.(\ref{b2}), there are two families of solutions. In limit $\alpha \rightarrow0$ for minus solution, we get $5D-LTB$ solution \cite{48}, however no results are available in literature if one apply limit $\alpha \rightarrow 0$ to plus branch solution, but such solutions are very interesting as these provide the contribution of pure Gauss-Bonnet terms, here we discuss the both these solutions and their singularity structure explicitly.
 \subsection{Minus Branch Solution}
 Maeda \cite{53}, found the LTB models near region $r \sim 0$, in the framework of EGB without finding an explicit solution. Also, Jhingan and Ghosh \cite{44} explored $5D-LTB-EGB$ gravitational collapse with dust matter as source of gravity. In this work, we determine the exact solution of $5D-LTB$ model in EGB gravity with \textit{perfect fluid} in the closed form to see the effect of pressure on the final fate of gravitational collapse.
 Hence, we consider minus branch solution with marginally bound case $Z=1$. Integrating Eq.(\ref{b2}), we have
\begin{eqnarray}\nonumber
 t_\varsigma(r)-t=\frac{\sqrt{\alpha}}{2\sqrt{2}} tan^{-1}\left[\frac{3C^2(1-\frac{4}{3} \alpha p_0) -\sqrt{C^4(1-\frac{4}{3} \alpha p_0) +8 \alpha \zeta}}{2\sqrt{2}C(1-\frac{4}{3} \alpha p_0)[\sqrt{C^4(1-\frac{4}{3} \alpha p_0)+8 \alpha \zeta}-C^2]^{1/2}}\right]\\\label{s3}+ \sqrt{\frac{\alpha C^2(1-\frac{4}{3} \alpha p_0)^2}{\sqrt{C^4(1-\frac{4}{3} \alpha p_0)+8 \alpha \zeta}-C^2}},
\end{eqnarray}
where $t_\varsigma (r)$ is function of integration which it is related to time formation of singularity.  Without the loss of generality, we consider  at $t=0$, $r$ coincide with the area radius
\begin{eqnarray}\label{s4}
 C(0,r)=r.
\end{eqnarray}
Above two equations lead to
\begin{eqnarray}\nonumber
t_\varsigma =\frac{\sqrt{\alpha}}{2\sqrt{2}}tan^{-1}[\frac{3(1-\frac{4}{3}\alpha p_0)-\sqrt{1-\frac{4}{3}\alpha p_0+8 \alpha \tilde{\zeta}}}{2 \sqrt{2}(1-\frac{4}{3}\alpha p_0)[\sqrt{1-\frac{4}{3}\alpha p_0+8 \alpha \tilde{\zeta}}-1]^{1/2}}]\\\label{a1}+\sqrt{\frac{\alpha (1-\frac{4}{3}\alpha p_0)^2}{\sqrt{1-\frac{4}{3}\alpha p_0+8 \alpha \tilde{\zeta}}-1}}  ,
\end{eqnarray}
where $\tilde{\zeta}=\zeta/r^4$. It is time formation of singularity which is effected by pressure and $\alpha$.

The Kretschmann scalar $K=R_{\mu\nu\gamma\lambda}R^{\mu\nu\gamma\lambda},$ given (\ref{met1}) with Eq.(\ref{s1}) takes the following form
\begin{eqnarray}\label{a1}
 K = 12 \frac{\ddot{C}^2}{C^2} +12 \frac{\dot{C^4}}{C^4} +4 \frac{\ddot{C'}^2}{C'^2}+12 \frac{\dot{C^2}\dot{C'^2}}{C'^2 C^2}.
\end{eqnarray}
It finite on the initial data surface. From field equations energy density is
\begin{eqnarray}\label{a1}
 \rho(t, r) = \frac{3 \zeta'}{2C^3 C'}-p_0.
\end{eqnarray}
Hence, it is clear that if $\zeta'$ is finite and strictly positive in the entire domain, then $\rho\rightarrow\infty$ when $C'=0$ and $C=0$. The shell crossing and shell focusing singularities occurs for $C'=0$ and $C=0$, respectively \cite{44}. Also, Kretschmann scalar diverges at $t=t_{c}(r)$, this implies the existence of curvature singularity \cite{29}.
 Using Eq.(\ref{s3}), shell focusing singularity curve is
\begin{eqnarray}\label{fs}
 t_{c}(r)=t_{\varsigma}(r)+\frac{\pi \sqrt{\alpha}}{4 \sqrt{2}},
\end{eqnarray}
The trapped surfaces are such surfaces whose outward normals are null
\begin{eqnarray}\label{h1}
 g^{\mu\nu} C_{,\mu}C_{,\nu} = - \dot{C^2}+\frac{C'^2}{B^2} = 0.
\end{eqnarray}
Now Eqs.(\ref{fv}) and (\ref{h1}) yield the horizon radius
\begin{eqnarray}\label{ha}
C(t_{AH}(r), r)= \frac{1}{\sqrt{p_0}}\sqrt{\sqrt{6p_0(\zeta -2 \alpha)+9}-3}.
\end{eqnarray}
In the above equation, the location of apparent horizons is effected by $\alpha$ and $p_0$.
Simplifying Eqs.(\ref{s3})and (\ref{fs}), we get horizon curve
\begin{eqnarray}\nonumber
 t_{c}(r)-t=\frac{\pi \sqrt{\alpha}}{4\sqrt{2}}+\sqrt{\frac{\alpha C^2(1-\frac{4}{3} \alpha p_0)^2}{\sqrt{C^4(1-\frac{4}{3}\alpha p_0)+8 \alpha \zeta}-C^2}}\\\label{25}+\frac{\sqrt{\alpha}}{2\sqrt{2}}tan^{-1}[\frac{3C^2(1-\frac{4}{3}\alpha p_0)-\sqrt{C^4(1-\frac{4}{3}\alpha p_0)+8 \alpha \zeta}}{2\sqrt{2}C(1-\frac{4}{3}\alpha p_0)[\sqrt{C^4(1-\frac{4}{3}\alpha p_0) +8 \alpha \zeta}-C^2]^{1/2}}].
\end{eqnarray}
Now combining Eqs.(\ref{ha}) and (\ref{25}), we get
\begin{eqnarray}\nonumber
 t_{c}(r)-t_{AH}(r)=\frac{\pi\sqrt{\alpha}}{4\sqrt{2}}+\sqrt{\frac{\alpha (Q-3)(1-\frac{4}{3}\alpha p_0)^2}{\sqrt{(Q-3)^2 (1-\frac{4}{3}\alpha p_0)+8 \alpha {p_0}^2 \zeta}-(Q-3)}}\\\label{a1}+\frac{\sqrt{\alpha}}{2\sqrt{2}}tan^{-1}[\frac{3(Q-3)(1-\frac{4}{3} \alpha p_0)-\sqrt{(Q-3)^2(1-\frac{4}{3} \alpha p_0)+8\alpha {p_0}^2 \zeta}}{2\sqrt{2 (Q-3)(1-\frac{4}{3} \alpha p_0)[(Q-3)^2(1-\frac{4}{3} \alpha p_0)+8\alpha {p_0}^2 \zeta -(Q-3)]^{1/2}}}],
\end{eqnarray}
where $Q=\sqrt{6p_0\zeta-12\alpha p_0+9}$.\\
It is to be noted that for $\alpha>0$, time difference between the formation of central singularity and apparent horizon is effected by the pressure. We would like to mentioned that all the results reduced to the dust case \cite{44}, when $p_0=0$.

\subsection{Plus Branch Solution}
The plus branch solution of Eq.(\ref{b2}) when subjected to marginally bound condition is given by
\begin{eqnarray}\nonumber
t_c(r)-t(r)&&=\frac{1}{2}log\mid8\alpha\zeta\mid-\frac{\sqrt{\alpha}}{2\sqrt{2}} log\mid C^{2}(1-\frac{4}{3}\alpha p_{0})\\\nonumber &&-\sqrt{C^{4}(1-\frac{4}{3}\alpha p_{0})+8\alpha \zeta}+\sqrt{2}C(1-\frac{4}{3}\alpha p_{0})(\sqrt{\sqrt{C^{4}(1-\frac{4}{3}\alpha p_{0})+8\alpha \zeta}+C^{2}})\mid\\&&+ \sqrt{\frac{\alpha C^2(1-\frac{4}{3} \alpha p)^2}{\sqrt{C^4(1-\frac{4}{3} \alpha p)+8 \alpha \zeta}-C^2}},
\end{eqnarray}
After applying the same procedure as in the previous case, we get

\begin{eqnarray}\nonumber
 t_c(r)-t_{AH}&&=\frac{1}{2}log\mid8\alpha\zeta\mid+\frac{\sqrt{\alpha}}{2\sqrt{2}} log\mid\frac{1}{p_{0}}\Bigg[(Q-3)\\\nonumber &&(1-\frac{4}{3}\alpha p_{0})+\sqrt{(Q-3)^{2}(1-\frac{4}{3}\alpha p_{0})+8\alpha \zeta}\\\nonumber
 &&+\sqrt{2(Q-3)}(1-\frac{4}{3}\alpha p_{0})(\sqrt{\sqrt{(Q-3)^{4}(1-\frac{4}{3}\alpha p_{0})+8\alpha \zeta}+(Q-3)})\Bigg]\mid\\&&
 + \sqrt{\frac{\alpha (Q-3)(1-\frac{4}{3} \alpha p_0)^2}{\sqrt{(Q-3)^{2}(1-\frac{4}{3} \alpha p_0)+8 \alpha \zeta}+(Q-3)}},
 \end{eqnarray}

This implies that horizons form after the formation of singularity, hence singularity is uncovered due to the absence of event horizons and end state of gravitational collapse is a naked singularity.

\section{Summary and Conclusion}
The Einstein Gauss-Bonnet theory of gravity is the low-energy limit of super symmetric string theory. Here, we have investigated the exact solution of the field equations in the frame work of EGB theory with LTB model which enclosed the inhomogeneous perfect fluid in $5D$. In order to do so, the marginally bound condition has been imposed on the dynamical equation. The conservation of energy-momentum tensor implies that $\frac {\partial p}{\partial r}=0,$ which produces the result $p=p(t).$ Further, it has been taken as constant $p_0$, by using some physical assumption as given by Eq.(\ref{44b}). The procedure along with marginally bound condition enables us to integrate the differential equation Eq.(\ref{f2}) analytically in the the closed form. It has been found that resulting solution implies the spherical inhomogeneous prefect fluid gravitational collapse. The coupling constant $\alpha$ has direct effect on resulting singularity and the end state of gravitational collapse is reversed. For $\alpha > 0$, there exist a naked singularity. The time formation of singularity and horizons is deeply effected by the pressure term. The presence of pressure also reduces the total matter density of the gravitating system.

The position of the apparent horizon in the spacetime is effected by the factor $2 \alpha$ and pressure $p_0$.
Due to the presence of second order curvature corrections in EGB gravity the out product collapse is a massive naked singularity, which is not admissible in $5D-LTB$. The predict about the regular initial data has been made which results to the formation of massive naked singularity that is contradiction to CCC. The singularity in this case is weaker as compared to the corresponding $5D-LTB$, therefore the existence of naked singularity in the present case is not a serious contradiction of CCC. According to Seifert conjecture \cite{14} when the strictly positive finite amount of matter undergoes to the gravitational contraction, then one point is always hidden, which is a counterexample to Seifert conjecture. It has been investigated that singularities are formed than horizons in the marginally bound case $(Z(r)=1)$, hence no BH is formed, and hence they must violate HC. In other words the present analysis is the counterexample to all three conjectures. But this analysis does not provide the serious threats to CCC.
\textbf{Following works with more general spacetimes are in progress in Gauss-Bonnet theory of gravity 
\begin{itemize}
\item The stability of gravitating source with heat flux and viscosity in Gauss-Bonnet theory of gravity 
\item The stability of charged gravitating source with heat flux and viscosity in Gauss-Bonnet theory of gravity 
\item Perfect fluid collapse with electromagnetic field in Gauss-Bonnet theory of gravity 
\item Tilted and non-tilted congruences in Gauss-Bonnet theory of gravity 
\end{itemize}}
\section*{Acknowledgment}
We appreciate the financial support from HEC, Islamabad, Pakistan under NRPU project with grant number 20-4059/NRPU/R \& D/HEC/14/1217. Also, we appreciate the constructive comments and suggestions of anonymous referee.
 
\end{document}